\documentclass[
secnumarabic,amssymb, nobibnotes, aps, preprint,superscriptaddress]{revtex4-1}

\usepackage[utf8]{inputenc}
\usepackage[T1]{fontenc}
\usepackage{bm}%
\usepackage{natbib}
\usepackage{wrapfig}
\usepackage{float}
\usepackage{graphicx}
\usepackage{multirow}
\usepackage{upgreek}
\usepackage{xcolor}

\begin{document}

\title{Heterogeneous cavitation from atomically smooth liquid-liquid interfaces}

\author{Patricia Pfeiffer}\email{patricia.pfeiffer@ovgu.de}
\affiliation{Institute of Physics, Otto-von-Guericke University Magdeburg, Universit\"atsplatz 2, 39106 Magdeburg, Germany}
\author{Meysam Shahrooz}
\affiliation{Dipartimento di Ingegneria Meccanica e Aerospaziale - DIMA, University of Rome ``Sapienza'', via Eudossiana 18, 00158 Roma, Italy}
\author{Marco Tortora}
\affiliation{Dipartimento di Ingegneria Meccanica e Aerospaziale - DIMA, University of Rome ``Sapienza'', via Eudossiana 18, 00158 Roma, Italy}
\author{Carlo Massimo Casciola}
\affiliation{Dipartimento di Ingegneria Meccanica e Aerospaziale - DIMA, University of Rome ``Sapienza'', via Eudossiana 18, 00158 Roma, Italy}
\author{Ryan Holman}
\affiliation{Image Guided Interventions Laboratory (GR-949), Faculty of Medicine, University of Geneva, Switzerland}
\author{Rares Salomir}
\affiliation{Image Guided Interventions Laboratory (GR-949), Faculty of Medicine, University of Geneva, Switzerland}
\affiliation{Radiology Department, University Hospitals of Geneva, Geneva, Switzerland}
\author{Simone Meloni}\email{mlnsmn@unife.it}
\affiliation{Dipartimento di Scienze Chimiche, Farmaceutiche e Agrarie - DOCPAS, University of Ferrara, via Luigi Borsari 46, 44121 Ferrara, Italy}
\author{Claus-Dieter Ohl}
\affiliation{Institute of Physics, Otto-von-Guericke University Magdeburg, Universit\"atsplatz 2, 39106 Magdeburg, Germany}

\date{\today}%

\maketitle

{\bf Pressure reduction in liquids may result in vaporization and bubble formation. This thermodynamic process is termed cavitation. It is commonly observed in hydraulic machinery, ship propellers, and even in medical therapy within the human body. While cavitation may be beneficial for the removal of malign tissue, yet in many cases it is unwanted due to its ability to erode nearly any material in close contact. Current understanding is that the origin of heterogeneous cavitation are nucleation sites where stable gas cavities reside, e.\,g., on contaminant particles, submerged surfaces or shell stabilized microscopic bubbles. Here, we present the finding of a so far unreported nucleation site, namely the atomically smooth interface between two immiscible liquids. The non-polar liquid of the two has a higher gas solubility and acts upon pressure reduction as a gas reservoir that accumulates at the interface. We describe experiments that clearly reveal the formation of cavitation on non-polar droplets in contact with water and elucidate the working mechanism that leads to the nucleation of gas pockets through simulations.}


Our understanding of the origin of bubble nucleation is still limited. The most accepted model requires pre-existing gas pockets stabilized in a hydrophobic pore \cite{Harvey1947,Crum1979,Atchley1989,Marschall2003,Borkent2009}. Once a sufficiently strong tension is applied, the gas expands explosively and forms a cavitation bubble. In typical systems these pores are provided by impurities or cracks on a submerged surface \cite{Holl1970,Crum1979}.

Only very few experiments \cite{angell1991,azouzi2013} demonstrated a cavitation threshold in water compatible with the classical nucleation theory \cite{Debenedetti1996}. Most experiments however suffer from a considerably lower threshold \cite{caupin2006} even when extreme care is taken in the preparation of the liquid. A possible explanation \cite{gao2021} might be nanoscale solid or gaseous nuclei \cite{Rossello2021} that reduce the threshold of an otherwise pure liquid.

Here, we report on a potentially novel source for cavitation: droplets of a highly non-polar liquid, which is hydrophobic and lipophobic, namely perfluorocarbons (PFC). The atomically smooth surface does not offer hydrophobic cracks for stabilization of pre-existing gas pockets. Liquid PFCs are very stable, chemically inert, feature high gas solubility, and used for \textit{in vivo} oxygen delivery and further biomedical applications \cite{ Riess2005, Lorton2018,holman2021}. The unique PFC properties are an effect of low self cohesion, polarity, and polarizability \cite{holman2021}. The PFC used in the experiments, i.\,e., 1-Bromoheptadecafluorooctane (PFOB), has an oxygen solubility that is about 20 times higher than water. In fact, pure PFOB theoretically dissolves 360 fold more oxygen than water in term of molar fraction and 25 times more in term of oxygen amount per unit of liquid volume  \cite{Riess2005,Lorton2018,Desgranges2019,holman2021}. This novel cavitation nucleus might be of interest for some medical applications, such as high intensity focused ultrasound ablation, localized drug delivery, and radiotherapy; or engineering technologies that could reduce cavitation in hydraulic machinery as pumps or injectors.

Here, we show the cavitation activity at a liquid-liquid interface of two immiscible fluids in a thin liquid gap. 
Rapid heating of the liquid through a laser-induced plasma generates a high pressure region that, besides nucleating a central bubble, also launches numerous waves in the solid and the liquid. A particularly interesting wave is the transverse Lamb-type wave that leads to strong tension in the liquid. This Lamb wave is sufficiently strong to nucleate cavitation bubbles within a gap containing water \cite{Rapet2020}. The same approach of strong tensile stress generation is used here, to study cavitation inception of PFC droplets. Atomistic simulations are performed on a water/PFC sample containing N$_2$ to mimic air dissolved in the liquids to investigate the cavitation origin in this system.



\begin{figure}
\centering
 \includegraphics[width=0.80\columnwidth]{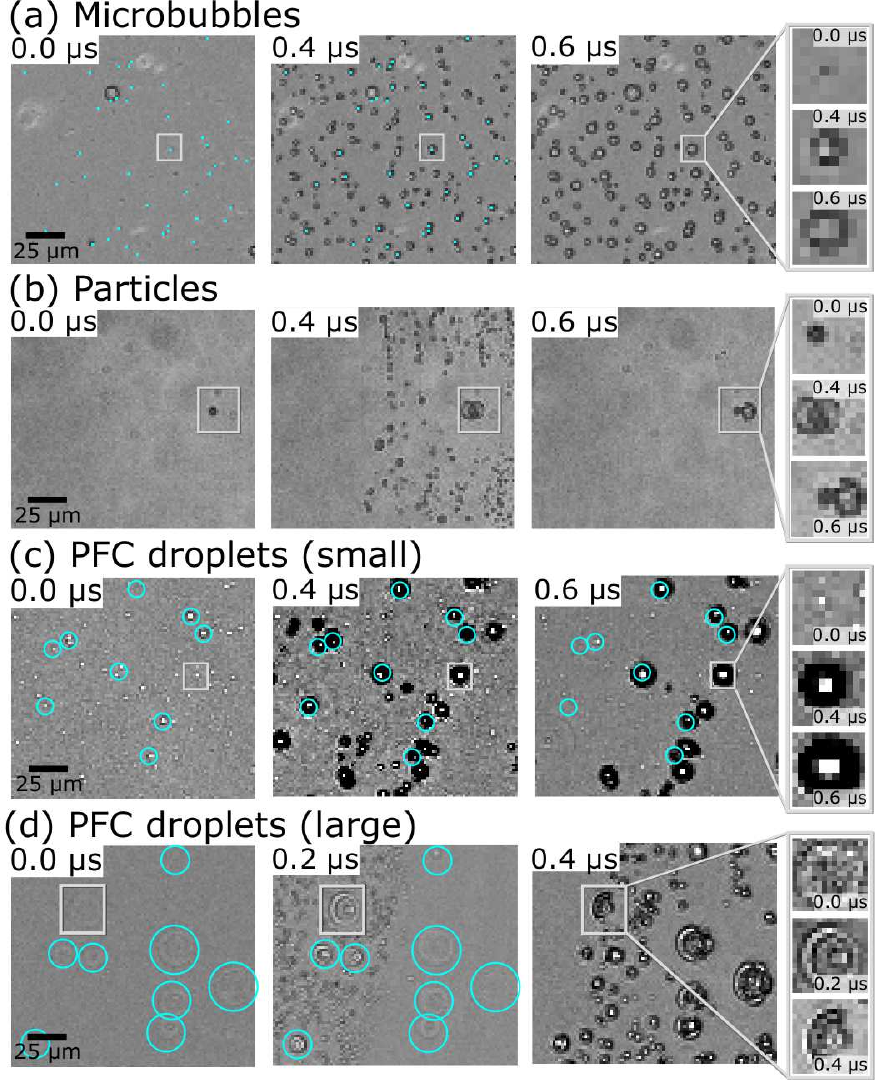}
\caption{Cavitation in a thin liquid gap. Experimental snapshots of secondary cavitation with different cavitation nuclei: (a) microbubbles, (b) magnetic beads, (c) and (d) PFC droplets before ($t=0$) and after the passage of the Lamb-type wave ($t=0.2 - 0.6\,\upmu$s). The cyan spots in (a) highlight some microbubbles, which expand after the passage of the wave. (b) A single magnetic bead (dark spot) forms two bubbles ($t=0.4\,\upmu$s) after the passage of the rarefaction wave. (c) The initial position of some PFOB droplets is marked with a cyan circle. After the wave passage bubbles are formed on these droplets. (d) On the 4H-PFOB droplets bubbles are formed mainly at the PFC/water interface. 
Scale bar: 25\,$\upmu$m.}
\label{fig:cav_nuclei}
\end{figure}

Figure \ref{fig:cav_nuclei} shows experimental snapshots of the cavitation activity in water loaded with (a) microbubbles, (b) particles and (c-d) PFC droplets. The primary cavitation is formed at the very left outside the field of view.

In (c-d) selected PFC droplets are marked with a cyan circle ($t=0$). When the Lamb wave passes, on most of these droplets a bubble emerges ($t=0.2-0.4\,\upmu$s), which will collapse after a short time and no visible bubbles remain. The PFC droplets are still intact and visible. The expansion of the primary cavitation drives a radial flow that transports the droplets from their original position to a slightly shifted one. During this transport some droplets may deform and eventually split up into several droplets.

Interestingly, a droplet may nucleate multiple times. Initially, some droplets nucleate a bubble upon tension. A later tension wave, which might be a result of reflections of waves within the glass, is able to nucleate bubbles at some of the PFC droplets previously acting as cavitation nuclei. This demonstrates that the droplets are not used up by one cavitation event but can serve multiple times as a cavitation nucleus (see Suppl. Information). This is called catalyst in chemistry. 

As the 1H,1H,2H,2H-Perfluorooctyl bromide (4H-PFOB) emulsions show larger droplets, we are able to see exactly where the bubbles are formed (Fig. \ref{fig:cav_nuclei}(d)). 
At $t=0.2\,\upmu$s secondary cavitation bubbles are formed in the left part of the frame. Here, pronounced circles are formed around the droplets. In the right part of the frame no bubbles are visible yet, as the Lamb wave has not entered this region. At $t=0.4\,\upmu$s the bubbles formed at the interface might break into smaller ones. Overall, we see that bubbles are mostly nucleated along the PFC/water interface.  

Atomistic simulations were run to investigate the origin of bubble nucleation. The computational sample, described in detail in the ``Methods'' Section, consisted of a PFC and a water slab in contact with each other. N$_2$ molecules were randomly inserted in the PFC slab to mimic air dissolved in the liquid. Additional nitrogen molecules were randomly inserted in the water slab as well. Given the very low solubility, most of the N$_2$ inserted in water drifted toward the PFC domain. This simple computational sample contains the key ingredient of the experimental sample, an interface between the two liquids where bubbles nucleate, though admittedly it lacks some more subtle features, such as a curved interface between the two liquids. The sample is thermalized at 1\,bar and 300\,K for 5\,ns and then is aged for 20\,ns during which we computed its properties. The Lamb-type wave is mimicked by subjecting the computational sample to a negative pressure of -20\,MPa. 

\begin{figure}
\centering
 \includegraphics[width=0.85\columnwidth]{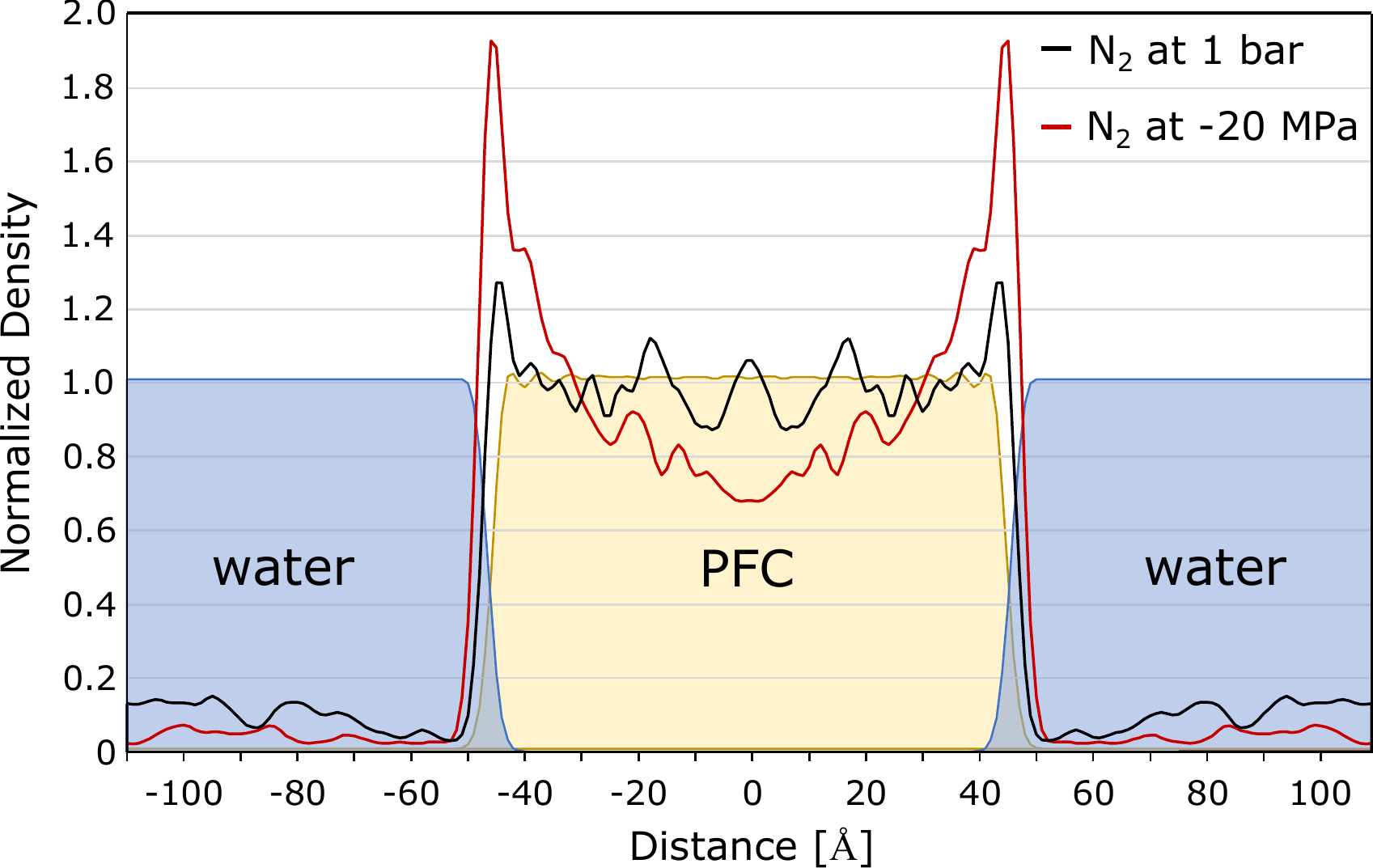}
\caption{Density profiles of the simulation sample. Density profile, normalized with respect to the bulk density value, of water (blue), PFC (yellow), and N$_2$ at 1\,bar (black) and -20\,MPa (red). Symmetry consistent with the computational sample, that the left and right PFC/water interfaces are equivalent, is enforced on the density profiles. 
The water and PFC densities drop smoothly at the interface between the two liquids. The PFC density shows the typical ripples observed for liquids at lyophobic interfaces. N$_2$ tends to accumulate at the interface already at 1\,bar, but this behavior is highly enhanced under tensile conditions mimicking the Lamb-type wave.}
\label{fig:sim}
\end{figure}

At ambient conditions, the sample shows an almost uniform density profile of N$_2$ in PFC, with a minor increase of concentration of the gas at the interface between the liquids; beyond the interface the N$_2$ density smoothly drops to a much lower density, as expected due to the much lower gas solubility in H$_2$O (Fig.~\ref{fig:sim}). For the same sample at -20\,MPa one observes a stark change in the N$_2$ density profile, with a significant reduction of density in the bulk domain of the PFC slab and a significant increase at the liquids' interface. The accumulation of the gas at the interface is accompanied by an increase of separation between the PFC and water slab. Considering the positions of the water and PFC as Gibbs surfaces, $x_{H_2O}$ and $x_{PFC}$, in the present case the position where the normalized water and PFC density is equal to $0.5$, the distance $\left | x_{H_2O} - x_{PFC} \right |$ between the two liquids grows from $\sim 0.65$\,\AA\ to $\sim 0.75$\,\AA. We believe that the accumulation of gas at the PFC/water interface follows a positive feedback mechanism: the decrease of pressure induced by the Lamb-type wave increases the separation between PFC and water, this makes room for air to accumulate in this emptier region and, in turn, the softer gas film forming at the interface helps to further separate the two liquids, making further space for accumulating additional air. 

When does the accumulation of $N_2$ (or $O_2$, air, etc.) at the PFC/water interface stop? The local increase of the gas at the interface between the liquids results in a corresponding decrease in the bulk PFC (and water), which in this process acts as a gas reservoir. This induces a decrease of the chemical potential of the gas in the liquid(s), which arrests the ever growing increase of concentration at the interface, hence the thickness of the film. Given the relatively small size of the computational sample, orders of magnitude smaller than the PFC droplets of the experiments,
the transfer of N$_2$ from the bulk of the PFC slab to the interface is arrested after few molecules moved. For the same reason, the growth of the distance $\left | x_{H_2O} - x_{PFC} \right |$ between the water and PFC surface under the action of mimicked Lamb wave ($-20$\,MPa pressure) is limited. We speculate that in the experimental sample, containing micrometer-sized droplets, the positive feedback mechanism described above is responsible for the air film observed around the PFC droplets. This film can result in the formation of bubbles according to two possible mechanisms, or their combination: i) The local gas supersaturation reduces the surface tension of the liquids consistent with results on bulk supersaturation of water~\cite{lubetkin2003much}, which reduces the bubble nucleation barrier~\cite{Debenedetti1996,giacomello2013geometry}, easing the process; ii) The gas film acts as a cavitation nucleus, which, perhaps due to inhomogeneities of the two liquids at their interface, can be destabilized, producing cavitation bubbles. 

Our findings have not only a fundamental impact in cavitation, but also implications in medicine, as PFC droplets are used as oxygen carrying blood substitutes, antihypoxants, and radiological contrast agents \cite{holman2021}. Currently, PFC emulsions are tested in preclinical high intensity focused ultrasound tumor ablation \cite{Kopechek2013,Moyer2015,Lorton2020,Desgranges2019}, as microbubbles have enhanced ablation and reduced treatment times \cite{Huang2019}. A potential advantage of droplets is the ability to be sonicated repeatedly without degradation, allowing lower applied powers, reduced prefocal interactions, and reduced treatment times. By understanding and exploiting the cavitation effects, the setup and sonication parameters might be optimized to improve local absorption and treatment efficacy for ablation therapy. 

Summarizing, experiments show that PFC droplets suspended in water may nucleate cavitation when exposed to sufficient tension, here through a Lamb-type wave. Simulations suggest that cavitation nucleation at the water/PFC interface is the result of a combination of the following key ingredients: i) the presence of a liquid with high air solubility, which acts as a gas reservoir, ii) the interface of this liquid with a second immiscible liquid, whose surfaces can be separated under the action of a Lamb-type wave, thus making room for the formation and accumulation of gas enriched film which, eventually, iii) results in the cavitation of a gas bubble either via reduction of the surface tension of water and/or PFC, or through the film acting as a cavitation nucleus. We believe that not only PFC droplets can induce cavitation, but any liquid immiscible with water that has also a high gas solubility. The investigation of the generality of the phenomenon described for the first time in this report is left for a forthcoming work.

\section*{Methods}
\subsection*{Experimental}
\begin{figure}
\centering
 \includegraphics[width=1\columnwidth]{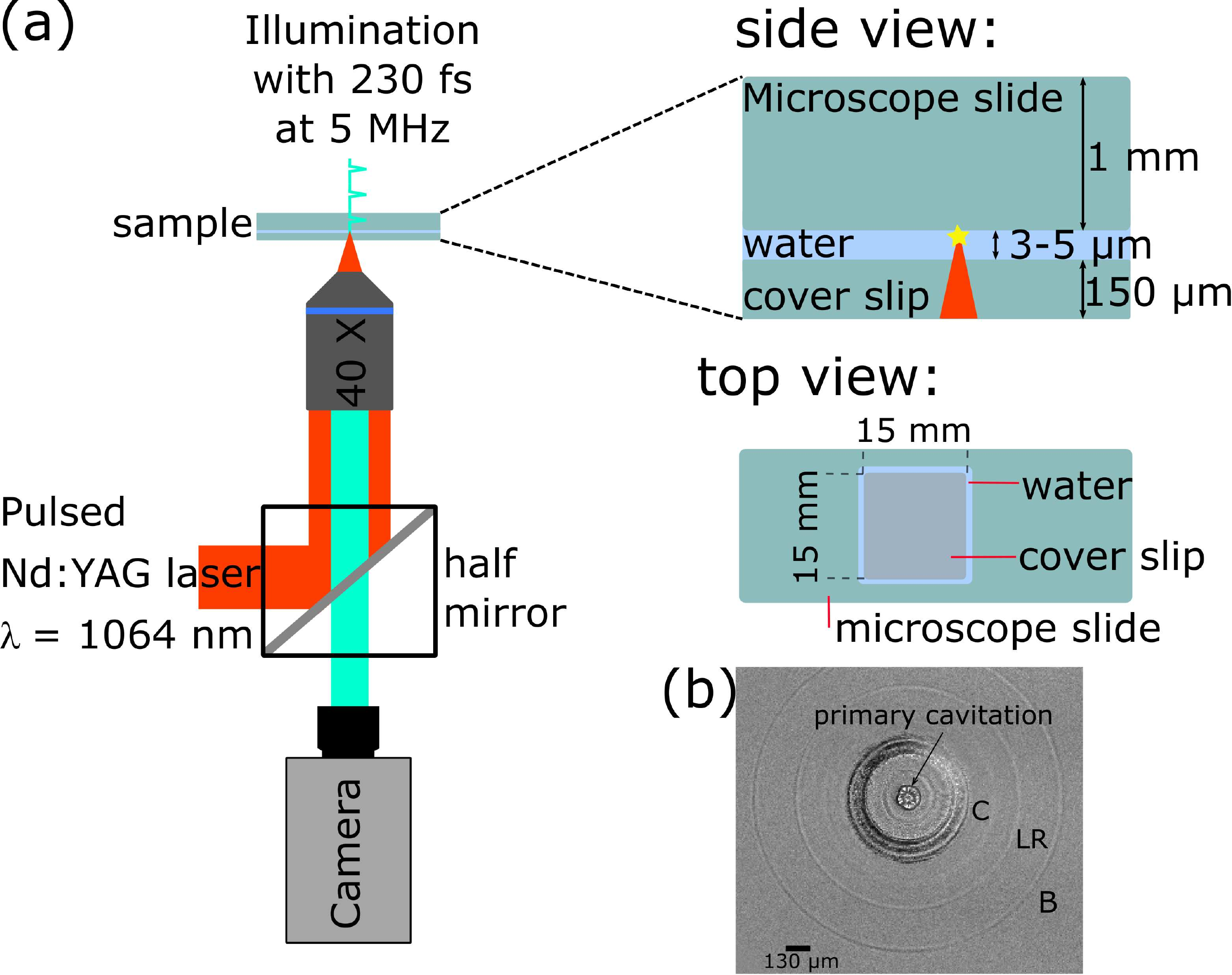}
\caption{Experimental setup for secondary cavitation inception and observation in a thin liquid gap. (a) The gap consists of two glass slides, sandwiching a $3-5\,\upmu$m thick liquid layer. (b) Waves occurring in the thin gap after laser-induced optical breakdown at $t=0.2\,\upmu$s. In the center the main cavitation bubble is formed (primary cavitation). B: longitudinal bulk wave, $v \approx 4000$\,m/s, LR: leaky Rayleigh wave $v \approx 2900\, m/s$, C: Lamb wave $v \approx 1450$\, m/s. The latter induces secondary cavitation (Lamb wave induced cavitation), which consists of many tiny bubbles formed in annular rings behind the wave.}
\label{fig:setup}
\end{figure}

High-speed video recording (HPV-X2, Shimadzu, Kyoto, Japan, $5 \times 10^6$ frames/s; spatial resolution 1.3 $\upmu$m per pixel) is used to observe the fast dynamics of cavitation inception in a thin liquid gap. Stroboscopic illumination is provided by a single light pulse in one frame from a femtosecond laser (FemtoLux 3 SH, EKSPLA, Vilnius, Lithuania, pulse duration 213\,fs at a maximum repetition rate of 5\,MHz, wavelength 515\,nm) (Fig. \ref{fig:setup}(a)).
The nanosecond laser pulse inducing the plasma appears $84\pm5$\,ns before a femtosecond laser pulse. The short illumination time enables us to visualize the waves traveling in the solid and in the liquid (Fig. \ref{fig:setup}(b)). The primary cavitation (central bubble) is surrounded by annular rings of cavitation bubbles that are created by the Lamb-wave (C). A leaky Rayleigh wave (LR) is considerably faster and travels in the solid along the surface. The longitudinal bulk wave (B) is the fastest wave traveling in the glass. 

Loading the gap with a fluorescent dye (Rhodamine 6G), the thickness can be measured with the help of a confocal microscope (SP8 Confocal Laser Scanning microscope, Leica GmbH, Wetzlar, Germany). Using a high magnification lens (HC PL FLUOTAR L 63x/0.70 DRY), a pixel resolution of 0.019 $\upmu$m/pixel was achieved. From the gray value profile, the inflection points were determined and from their distance the thickness of the gap was calculated, having a typical value of $3.4 \pm 0.1\, \upmu$m.

Different systems are studied: microbubbles ("Luminity", \textit{CS Diagnostics GmbH}; ultrasonic contrast agent, which is perflutren coated with a lipid layer), particles (magnetic beads, tosyl activated in liquid solution; provided by M. Pumera, UCT Prague Nanorobots Research Center), and PFC droplets in water with 1-Bromoheptadecafluorooctane (PFOB; 99\%; purchased from \textit{Sigma-Aldrich}), and 1H,1H,2H,2H-Perfluorooctyl bromide (4H-PFOB; 97\%, purchased from \textit{Fluorochem}, UK). The emulsions are produced by mixing 1\,\% PFC with water. The mixture is sonicated for 5\,min (0.5\,Hz) with a Bandelin Sonopuls Ultrasonics Homogenizer (UW 2200 with the titanium tip MS72) at a power of 36\,W. The PFOB droplets have a mean diameter of 2-3\,$\upmu$m, whereas the 4H-PFOB emulsion shows a droplet size between 6-25\,$\upmu$m. 

A droplet (10\,$\upmu$l) of the sample is placed on a microscope slide (\textit{Paul Marienfeld GmbH \& Co. KG}, Germany) and covered carefully with a cover slip (\textit{Menzel Gl\"aser} \#1, Germany), such that no air bubbles remain in the gap. Both glass slides are clamped together. To avoid drying out of the gap which could affect the thickness of the liquid layer an additional droplet of liquid is placed close to the gap.

\subsection*{Simulation Details}
\label{sec:ComDetails}

\begin{figure}[ht]
	\centering
	\includegraphics[width=1\columnwidth]{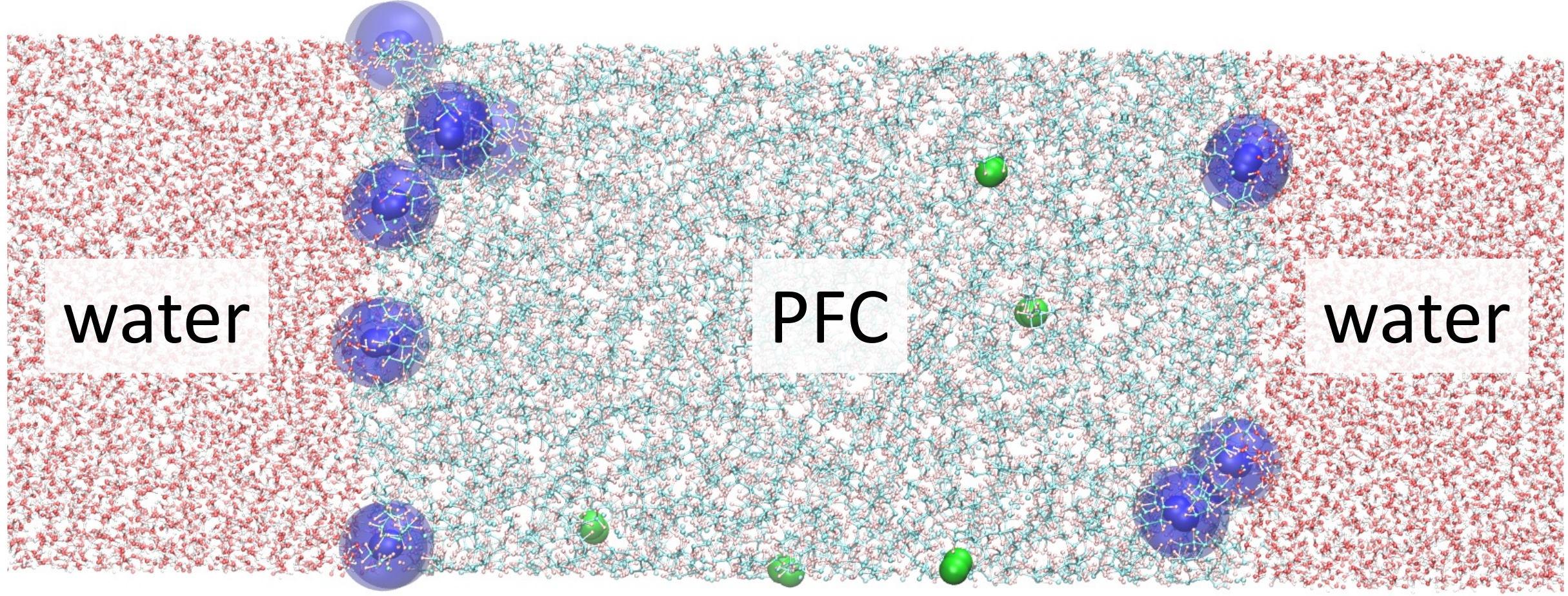}
	\caption{
	 Snapshot of the computational sample at -20\,MPa. The sample consists of a slab of PFC and a slab of water (the two halves in the pictures are connected through periodic boundary conditions). To highlight N$_2$, the atoms of these molecules are represented as spheres radius equal to the van der Waals radius of nitrogen, while the liquids are shown in a ball-and-stick representation. Thus, the size of spheres representing nitrogen, water and PFC do not represent the actual size of the various molecules. To further highlight N$_2$, blue spheres are drawn around them. Consistently with the N$_2$ density of Fig. \ref{fig:sim}, this snapshot shows an accumulation of gas at the two water-PFC interfaces.}
	\label{fig:simbox}
\end{figure}

The computational sample (Fig. \ref{fig:sim}) consists of a slab of 770 PFC molecules and a slab of $\sim$ 15500 water molecules, for a total of $\sim$ 66500 atoms; 14 nitrogen molecules are introduced randomly inside the PFC slab using Packmol \cite{Martinez2009} to model the dissolved air. 
The N$_2$ molar fraction in our computational sample is approximately five times higher than the experimental value~\cite{Battino1984}. This higher N$_2$ concentration allows us to determine the properties of the dissolved gas with an adequate statistical accuracy within the timescale accessible by molecular dynamics. In particular, this oversaturation allows one to produce an appreciable accumulation of gas at the interface under the effect of the simulated Lamb-type wave. Costa Gomes \emph{et al.}~\cite{CostaGomes2007} have already successfully used this approach to investigate the properties of O$_2$ dissolved in PFC using a gas concentration $\sim 12.5$ higher than the (computational) solubility of the gas. We validated this approach considering bulk PFC/N$_2$ samples at growing gas concentration and have verified that the structural characteristics of the mixture, e.\,g., the partial pair correlation functions of N$_2$ with PFC and with other gas molecules (see Suppl. Information) do not significantly change among the experimental saturation concentration and the one used in our simulations of the interface system. The suitability of our simulations approach is also confirmed by the uniform distribution of N$_2$ in the bulk of PFC at $1$\,bar: despite the higher concentration, no gas bubbles are formed in the PFC bulk along the duration of the simulation (Fig.~\ref{fig:simbox}).

MD simulations were run at constant temperature (300\,K) and pressure (1\,bar and -20\,MPa) using the Nos\'e-Hoover chain thermostat~\cite{Martyna1992} and the Martyna-Berne-Klein barostat~\cite{Martyna1994}. The simulation box is allowed to expand and compress only along the direction orthogonal to the interface, which corresponds to the $x$-direction in the chosen reference frame. The 1D constant pressure algorithm is applied to prevent the nonphysical shrinking of the simulation box on the PFC/water $y-z$ surface plane resulting from the tendency of the flat interface system to minimize the surface area between the two liquids. Indeed, in a genuine 3D system PFC forms spherical bubbles, as seen in the experiments, and the water/oil interface cannot shrink under the action of the interface energy. It is worth remarking that a genuine 3D computational system containing a spherical droplet of a size  treatable by extensive molecular dynamics (one or few nanometers in diameter) will probably introduce severe artifacts. For example, in nanometer size PFC droplets the Laplace pressure is much higher than in the micrometer size experimental counterpart. Summarizing, the flat interface system with variable cell along the $x$ direction, the direction orthogonal to the PFC/water interface, is the best compromise between accuracy and feasibility of simulations.

The sample was initially thermalized for 5\,ns at 300\,K and 1\,bar; after thermalization, simulations were run for additional 20\,ns, during which we collected data to analyze the gas density distribution. In parallel, the 300\,K / 1\,bar thermalized sample was brought to -20\,MPa, thermalized for an additional 5\,ns, and then evolved for an additional 20\,ns for analysis.

Forces acting on atoms are derived from TIP4P/2005 model for water \cite{Abascal2006} and the GAFF force field for PFC. N$_2$ is modeled as a rigid molecule. Molecules of the different species interact via electrostatics and van der Waals forces, the latter modeled by the Lennard-Jones potential $\nu \! \left( r_{ij} \right) = 4 \varepsilon_{\alpha\beta} \left[ \left( \sigma_{\alpha\beta}/r_{ij} \right)^{12}) - \left( \sigma_{\alpha\beta}/r_{ij} \right)^{6} \right]$, where $r_{ij}$ represents the distance between two atoms belonging to different molecules, and $\varepsilon_{\alpha\beta}$ and $\sigma_{\alpha\beta}$ are the characteristic energy and distance, respectively, between atoms of type $\alpha$ and $\beta$. As customary, cross-species coefficients, i.\,e., those between atoms of different types, are obtained from same-species parameters through the Lorentz-Berthelot combination rules, $\varepsilon_{\alpha\beta} = \sqrt{\varepsilon_{\alpha}\varepsilon_{\beta}}$ and $\sigma_{\alpha\beta} = \left( \sigma_{\alpha} + \sigma_{\beta} \right) \! /2$. N$_2$ parameters for the cross-species coefficient have been taken from Ref.~\cite{jiang2005separation}.

\section*{Data availability}
The authors declare that data supporting the findings of this study are
available within the paper and its supplementary information.
Additional data that support the findings of this study are available from the authors on reasonable request.

\section*{Code availability}
Codes used for present simulations are either community codes available to any reader free of charges, namely the LAMMPS molecular dynamics software, or analysis codes to compute and symmetrize the density profiles, provided in the supplementary information.

\providecommand{\noopsort}[1]{}\providecommand{\singleletter}[1]{#1}%

\section*{Acknowledgments}
M. Pumera is acknowledged for providing the magnetic beads. We thank A. Eremin for the confocal thickness measurements.
This work was financially supported by the European Social Fund (No. ZS/2019/10/103050) as part of the initiative “Sachsen-Anhalt WISSENSCHAFT Spitzenforschung/Synergien”.
This study was partially supported by ``Fondo per l’incentivazione alla ricerca (FIR), 2020'' from the University of Ferrara in Italy.

\section*{Author Contributions}

C.-D. O. designed the study, R. S. selected the inclusion, P. P. performed the experiments and analyzed the data; S. M. designed the simulation campaign; M. S. performed the simulations with the help of M. T.; S. M., M. S., M. T. and C. M. C. analyzed the simulation data; P. P. and S. M. and R. H. wrote the paper. All authors discussed the results, read, revised, and approved the final version.

\section*{Competing Interests}
The authors declare no competing interests.

\end{document}